\documentclass[a4paper,11pt]{article}

\usepackage[nosort,compress]{cite}

\usepackage{pos}

\title{Decimation map in 2D for accelerating HMC}

\author*[a,b]{Nobuyuki Matsumoto}
\author[a]{Richard C. Brower}
\author[c,b]{Taku Izubuchi}

\affiliation[a]{
  Boston University, Boston, MA 02215
}
\affiliation[b]{
  RIKEN/BNL Research Center, Upton, NY 11973, USA
}
\affiliation[c]{
  Brookhaven National Laboratory, Upton, NY 11973, USA
}

\emailAdd{nmatsum@bu.edu}

\abstract{
  To accelerate the HMC with field transformation,
  we consider a variant of the trivializing map,
  the decimation map,
  which can be regarded as a coarse-graining transformation.
  Using the 2D $U(1)$ pure gauge model,
  combined with the guided Monte Carlo algorithm,
  we show that the integrated autocorrelation time
  of the topological charge can be exponentially improved
  in the wall clock time.
  Our study indicates that
  incorporating renormalization group picture is
  a powerful and essential ingredient
  to accelerate the HMC at large $\beta$.
}

\FullConference{The 40th International Symposium on Lattice Field Theory (Lattice 2023)\\
  July 31st - August 4th, 2023\\
  Fermi National Accelerator Laboratory\\}

\begin{document}
\maketitle

\section{Introduction}
\label{sec:intro}

Critical slowing down
is intrinsic to Monte Carlo study of lattice quantum field theories,
but is not yet resolved satisfactorily.
The state of the art algorithm for the gauge field generation in QCD
is the Hybrid Monte Carlo (HMC) \cite{Duane:1987de},
and there are two major directions to cope with the
critical slowing down in this algorithmic framework
(see \cite{Gurtej2023, Boyle2023} for reviews).
One is
to align the velocity in the molecular dynamics (MD)
among all the Fourier modes
\cite{Parisi:1984,Batrouni:1985jn,Davies:1987vs,Katz:1987ti,Davies:1989vh},
and the other is to construct a field transformation
such that the resulting effective action
has advantageous sampling properties
\cite{Luscher:2009eq, Engel:2011re}
(see also \cite{Albergo:2019eim,Foreman:2021ixr,Albandea:2021lvl,Foreman:2021ljl,Caselle:2022acb,Bacchio:2022vje,Boyle:2022xor}).

This study follows the latter approach
following L\"uscher's seminal work \cite{Luscher:2009eq},
and we consider trivialization of link variables.
A difference, however, is that
we divide the lattice into local blocks
while the original work considered the global trivialization.
In fact, our field transformation
corresponds to eliminating links from the theory,
namely {\it decimation} of the variables.
When a link is trivialized,
the effective action for the surrounding links
exactly agrees with that of the integrated theory.
Our approach thus incorporates
the idea of renormalization group.

In general, such an algorithm can be extremely complex.
We thus first test our idea with the simplest gauge theory possible:
the pure $U(1)$ gauge theory in two spacetime dimensions.
The peculiarity of this model is that
there is no propagation mode;
the correlation length defined by the plaquette correlation function is zero.
A nontrivial quantity is the
topological charge under the periodic boundary condition,
which counts the winding number of
$U(1)$ plaquette angles around the faces,
{\it i.e.}, the number of vortices.
The density of vortices is determined by the susceptibility,
which is finite in the continuum limit.
In other words,
there is a typical volume scale in which a vortex appears.
Accordingly,
we need to move the variables collectively
on fine lattices to induce topological tunneling.
In fact, in the HMC,
we observe that the tunneling becomes exponentially hard
as in QCD.
The goal of this study is to demonstrate
that the decimation map can reduce the
autocorrelation of the topological charge significantly.

The algorithm consists of two parts
as in the original trivializing map.
We first construct
a series of local trivializing maps,
namely the {\it decimation map},
by solving the linear equation
for the gradient flow kernel.
We here parameterize the function space with Wilson loops
and directly invert the linear equation with an iterative solver.
We then use the guided Monte Carlo \cite{Horowitz:1991rr} for configuration generation,
which is a variant of the HMC
replacing the action in the MD Hamiltonian
by an approximate one.
The effects of finite flow step size
and the approximated action
are under control
and the volume scaling is power-law
while keeping the exact detailed balance.

We test our algorithm on a small system of
the physical volume $V_{\rm phys} = 6^2/g^2$,
where $g$ is the dimensionful coupling constant.
This volume corresponds to the typical scale of a single vortex.
We show that
we obtain $\times 73$ speedup
in the integrated autocorrelation time in wall clock
at $\beta=7.1$ ($16 \times 16$ lattice),
and furthermore,
the exponent towards the continuum limit is decreased
with a factor of $0.62$.
Our decimation map
exemplifies that
the HMC of a gauge theory can be accelerated
with a suitable field transformation.

\section{Decimation map}
\label{sec:decimation_map}

\begin{figure}[thb]
  \centering
  \includegraphics[width=150mm]{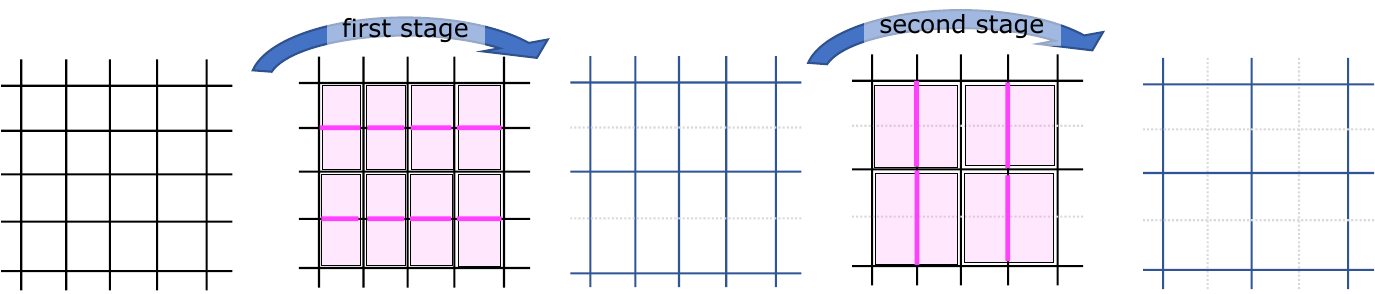}
  \caption{The basic idea of
    the decimation map is a successive local trivialization of link variables. At each stage,
    a set of independent link variables (colored with magenta) are chosen and trivialized.}
  \label{fig:decimation}
\end{figure}

The idea of
the decimation map can be described simply in the 2D $U(1)$ model
with the Wilson action:
\begin{align}
  S(U) \equiv -\beta \sum_x \cos \kappa_x,
\end{align}
where $\beta = 1/(ag)^2$ with
the lattice spacing $a$
and the coupling $g$
that determines the scale in the system.
$\kappa_x$ is the plaquette angle:
\begin{align}
  \kappa_x
  \equiv
  \frac{1}{i}
  \log
  (U_{x,0}U_{x+0,1}U_{x+1,0}^\dagger U_{x,1}^\dagger).
  \label{eq:plaq_angle}
\end{align}
As in figure~\ref{fig:decimation},
we iteratively choose
a set of independent link variables to trivialize.
Each stage of trivialization
doubles the size of
fundamental Wilson loops in the theory.

Let us write by $U'_{x,\mu}$ the link variables to be trivialized and
by $\tilde U_{x,\mu}$ the remaining variables.
Once the trivialization is performed successfully with the map:
\begin{align}
  U'_{x,\mu} = {\cal F}_{x,\mu}(V; \tilde U),
\end{align}
where $V_{x,\mu}$ are the trivialized variables,
the action will be transformed accordingly
from the original one $S=S(U',\tilde U)$ to:
\begin{align}
  S_{\rm eff} (\tilde U)
  \equiv
  S(U'(V; \tilde U),\tilde U)
  -
  \ln \det {\cal F}^*(V; \tilde U),
\end{align}
where ${\cal F}^*$ is the Jacobian of the map ${\cal F}$.
Then one can immediately see that $S_{\rm eff}$
agrees with the action we obtain
after integrating over $U'$
from the original action:
\begin{align}
  \int (dU')(d\tilde U)
  e^{-S(U', \tilde U)}
  =
  \int (d\tilde U)
  e^{-S_{\rm eff}(\tilde U)}.
\end{align}
We thus see that $S_{\rm eff} (\tilde U)$
describes a coarse-grained system
and each stage of the map corresponds to
the decimation of links.

\begin{figure}[htb]
  \centering
  \includegraphics[width=80mm]{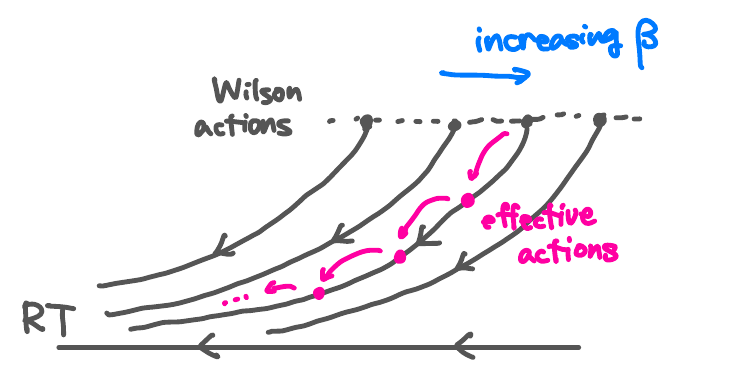}
  \caption{A qualitative picture describing the resulting flow in the action space.
    It is likely
    that the theory approaches towards the renormalized trajectory (RT),
    observing that the tunneling rate is not as high as
    for the Wilson actions of the larger lattice spacings,
    at least for the first a few stages.
    However, it is uncertain how far this picture holds in the flow
    and its precise structure is out of the scope of this work.}
  \label{fig:rt}
\end{figure}

Such observation has an important implication
that the decimation map
drives the system away from the continuum limit,
where the HMC becomes inefficient
(see figure~\ref{fig:rt}).
One can therefore expect that HMC in the decimated system
has smaller autocorrelation than in the original system.
After obtaining configurations
in the decimated system,
one can use ${\cal F}$ to supply (or {\it integrate in})
the decimated variables to calculate observables
for the original action.

\section{Obtaining the map numerically}
\label{sec:obtaining_map_numerically}

In our simple example,
there are only two Wilson loops attached to each link $U_{x,\mu}$.
Let us write by $\kappa_{1,2}$ the two phase angles
of the loops.
One can take the basis
for the function space to trivialize $U_{x,\mu}$ as:
\begin{align}
  e_{m_1, m_2} \equiv \cos (m_1 \kappa_1 + m_2 \kappa_2)
  \quad
  (m_1, m_2 \in \mathbb{Z}).
  \label{eq:basis}
\end{align}
Note that when $m_1=m_2$, they do not depend on $U_{x,\mu}$,
and thus $e_{m,m}$ $(m\in \mathbb{Z})$
are the zero modes of the differential operators.

As in \cite{Luscher:2009eq}, we take the gradient flow ansatz:
\begin{align}
  \dot U_{t, x,\mu} U_{t, x,\mu}^{-1}
  =
  \partial K_t,
  \quad
  U_{t, x,\mu} \equiv {\cal F}_{t,x,\mu}(V),
  \quad
  {\cal F}_{t=1}={\cal F},
\end{align}
where $\partial$ is the derivative with respect to
$\phi \equiv \log U_{x,\mu}$.
Then, one can write down the equation for $K_t$
by demanding the effective action to decrease linearly in $t$.
For example,
for the first stage:
\begin{align}
  [-\partial^2 + \beta t (\sin\kappa_1 - \sin \kappa_2) \partial ] K_t
  =
  \beta
  (\cos \kappa_1 + \cos \kappa_2) + \sum_m e_{m,m} v_{t,m},
  \label{eq:linear_equation}
\end{align}
For the higher stages,
terms with various $m_i$ appear on both hand sides.
By representing the differential operator in the basis~\eqref{eq:basis},
one can numerically solve the linear equation~\eqref{eq:linear_equation}
with, {\it e.g.}, CGNE with preconditioning by $\partial^{-2}$
in the space without the zero modes.
The zero mode part $\sum_m e_{m,m} v_{t,m}$ can be obtained
acting the differential operator to the solution
after solving the equation in the projected space.
We find that for $\beta \lesssim 8.9$,
the range $|m_{1,2}| \leq 128$ is sufficient
(see figure~\ref{fig:flow_coeffs}).
The inversion takes a few minutes to about a day using GPU
depending on the value of $\beta$ and the stage of decimation.
The details of the inversion will be described
in the subsequent paper.

In practice, we discretize the flow with the Runge-Kutta methods.
We here choose the midpoint integrator. 
For the map to be bijective,
we ensure that the step size is kept within a bound
that can be derived as in \cite{Luscher:2009eq}.
For the midpoint integrator,
given a kernel of the form:
\begin{align}
  K_t(\phi) = \sum_{m_1, m_2} e_{m_1, m_2} c_{t,m_1,m_2},
  \label{eq:flow_kernel}
\end{align}
the step size $\epsilon$ must satisfy:
\begin{align}
  \epsilon
  B_{t+\epsilon/2}\cdot
  \left(
  1+ \frac{\epsilon B_t}{2}
  \right) < 1,
  \quad
  B_t \equiv \sum_{m_1,m_2} (m_1-m_2)^2 |c_{t,m_1,m_2}|.
  \label{eq:bijection}
\end{align}

\section{Guided Monte Carlo}
\label{sec:guided_mc}

The functions $e_{m,m}$ are the Wilson loops of the double size
winding $m$ times.
They are independent of $U_{x,\mu}$,
but have different dependencies on the surrounding links.
They are the basis functions for determining the weights in the transformed system.
In fact, the integral of the zero mode over $t$
gives the next effective action:
\begin{align}
  S_{\rm eff} = \sum_m e_{m,m} \int dt\, v_{t,m}.
  \label{eq:approximation}
\end{align}
One can use this knowledge to simplify the force calculation
with the help of the guided Monte Carlo algorithm \cite{Horowitz:1991rr}
as below.

The key point of the guided Monte Carlo algorithm is that,
as far as the time evolution in the HMC is symplectic,
the detailed balance is exact.
However, since the energy conservation is used to
have nonvanishing acceptance in the Metropolis test,
it is important that
the Hamiltonian associated to the
symplectic integrator is close to the exact one.

We replace the action in the MD Hamiltonian
by an approximate one
to evade
the evaluation of the gradient of the Jacobian
as well as the force propagation.
Since the zero modes $v_{t,m}$
are calculable at each flow time $t$
as a sum of Wilson loops,
we can calculate the effective action
directly in the loop space
using eq.~\eqref{eq:approximation}.
We use the Simpson's formula
for approximating the right hand side,
which involves only one additional evaluation of $K_t$ at
$t=0$ to those calculated for the midpoint integrator.
The difference between
this approximate action
and the effective action of the discretized flow
comes from the error in the midpoint integrator and is $O(\epsilon^2)$,
where $\epsilon$ is the step size of the flow.
Note that, though for a fixed $\epsilon$
the acceptance rate becomes exponentially small
when enlarging the volume,
this effect can be compensated by
decreasing the step size $\epsilon$ of the trivializing flow
just as decreasing the MD step size in the conventional HMC.
The cost scaling in enlarging the volume
is therefore power law.

We set the approximate effective action
in the trivialized region to be constant in the MD,
and thus we choose the inner links purely randomly.
This corresponds to assigning the zero mass in the MD.
Nontrivial updates are performed for
the outer remaining links
through the Wilson loops in the large unit.
In the Metropolis test, however,
since the replacement involves discretization error of the flow,
we need to include all the updated links in the weight function.

In practice,
we choose the step size adaptively
by making it proportional to the bijection bound~\eqref{eq:bijection}
while the overall scaling is determined to keep
the acceptance around 0.8.
The adaptively chosen step size becomes
extremely small at large $\beta$,
however,
for the following two reasons.
One is that
the number of relevant basis functions
grows rapidly as $\beta$ is increased,
which will in turn decrease the bijection bound~\eqref{eq:bijection}
(see also figure~\ref{fig:exe_time} on this rapid growth).
The other is to control the acceptance rate
in the guided Monte Carlo for increasing the lattice sites.
In our calculation,
the second point is the bottleneck.
This cost due to the latter point may be circumvented
by using a higher-order integrator
or by switching to calculating the exact force of the effective action
as in \cite{Luscher:2009eq}.
As will be shown in section~\ref{sec:results},
the exponent of the cost towards the continuum limit can be decreased
with the decimation map,
and thus one can expect an acceleration with the latter algorithm.
However, this point needs to be verified with an additional study.

The separation of the inner and outer links
corresponds to separating the UV and IR modes in the system.
For simplicity, suppose that the decimation map is obtained without discretization error.
One can then choose the inner links without autocorrelation.
However, since the environmental outer variables are fixed,
this fluctuation is only around a fixed background configuration
and depends only on the local links
surrounding the trivialized region.
It can be thus interpreted as the UV fluctuation.
On the other hand, the update of the outer links is
completely insensible to the inner links,
whose dynamics is determined solely by the coarse-grained action.
Thus, the global update of the outer links corresponds to updating the IR modes.
By using this separation,
we can investigate which part of the algorithm is responsible
for accelerating the algorithm.

\section{Results}
\label{sec:results}

To enumerate the effectiveness of our approach,
we calculate the integrated autocorrelation time:
\begin{align}
  \tau_{\rm int}({\cal O}) \equiv
  \frac{1}{2}
  +
  \sum_{i \geq 1} \frac{ \langle {\cal O}_0 {\cal O}_i \rangle}
  {\langle {\cal O}_0 {\cal O}_0 \rangle},
\end{align}
where ${\cal O}_i \equiv {\cal O}(U_i)$
is the value of the observable ${\cal O}$
evaluated with the $i$-th configuration $U_i$.
Our interest is in the topological charge \cite{Luscher:1981zq, Phillips:1985sgx}:
\begin{align}
  Q \equiv -\frac{1}{2\pi} \sum_x \kappa_x,
\end{align}
which is integer-valued.
It is assumed to take the principal branch of the logarithm
to evaluate $\kappa_x$ in eq.~\eqref{eq:plaq_angle}.
One can derive analytically that
the susceptibility approaches in the continuum limit to:
\begin{align}
  \chi_Q
  \equiv
  \frac{\langle Q^2 \rangle}{ V_{\rm phys } }
  \to \frac{g^2}{(2\pi)^2}
  \quad
  (a\to 0),
\end{align}
which shows that a vortex (an instanton)
appears typically in a volume of $(2\pi)^2/g^2$.

We fix the physical volume to $V_{\rm phys} = 6^2/g^2$,
which is about the scale of a vortex,
and take the continuum limit $a\to 0$ by fixing $g$.
The trajectory length of the MD is
fixed to $1.0$ in all simulations.
In the conventional HMC,
the MD step size is scaled
such that the acceptance rate is around 0.8.
In the guided Hamiltonian simulations,
the step size is fixed to 0.05 ({\it i.e.,} 20 steps);
though this is not optimal,
this number is not relevant to the actual cost
because the algorithmic overhead is
in the field transformation part.
As mentioned in section~\ref{sec:guided_mc},
flow step size of the decimation map
determines the acceptance rate,
which is scaled
for the acceptance to be around 0.8.

\begin{figure}[htb]
  \centering
  \includegraphics[width=75mm]{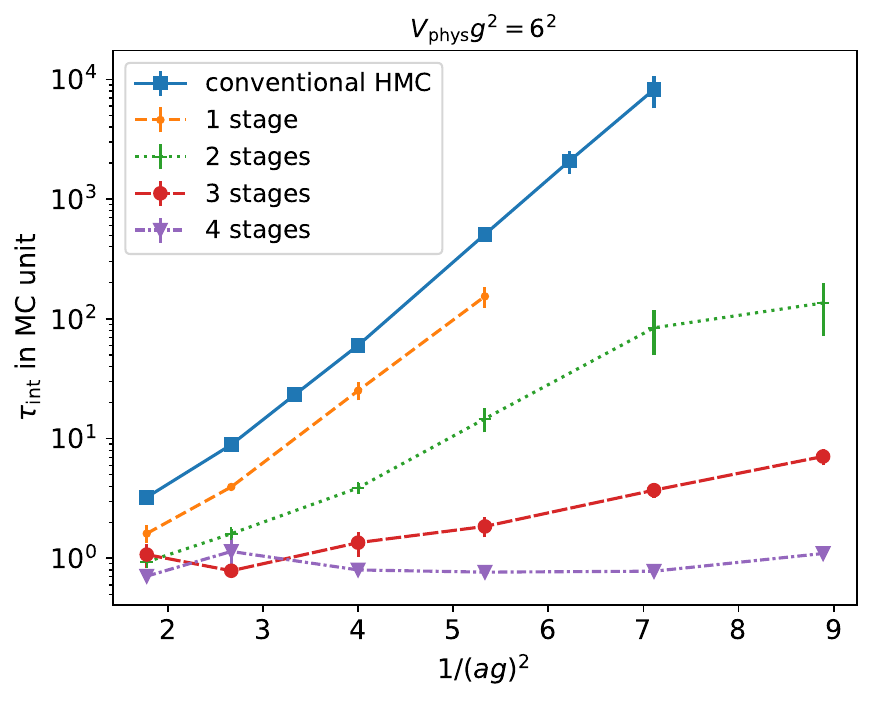}
  \includegraphics[width=75mm]{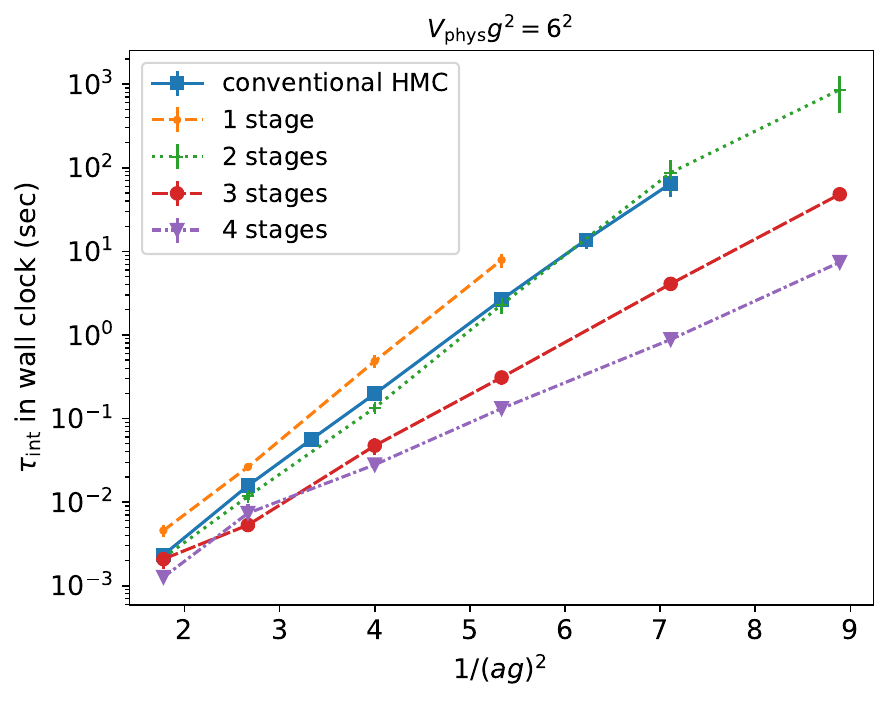}
  \caption{$\tau_{\rm int}(Q)$ in Monte Carlo steps (left)
    and in wall clock time (right).
    The conventional HMC (blue solid line)
    is compared to the guided Monte Carlo with
    the decimation map
    up to four stages.
    In Monte Carlo step,
    $\tau_{\rm int}(Q) \simeq 8000$
    at $\beta=7.1$ ($16\times 16$ lattice)
    is reduced to
    $\tau_{\rm int}(Q)\simeq 1$.
    In wall clock,
    the speedup at $\beta=7.1$ is x73
    and the decrease of the exponent
    is by a factor 0.62 with four stages.
  }
  \label{fig:tauint}
\end{figure}

Figure~\ref{fig:tauint} shows the scaling of
$\tau_{\rm int}(Q)$ in the units of
Monte Carlo steps and wall clock time.
The number of stages is taken up to four.
The computation is performed on CPU without parallelization.
As shown in the left panel,
for $\beta=7.1$ ($16\times 16$ lattice),
$\tau_{\rm int}(Q) \simeq 8000$ with the conventional HMC
is reduced to
$\tau_{\rm int}(Q)\simeq 1$
with the decimation map in the Monte Carlo unit.
In wall clock,
we see x73 speedup at $\beta=7.1$,
and furthermore,
the exponent has decreased with a factor of 0.62.

\begin{figure}[htb]
  \centering
  \includegraphics[width=70mm]{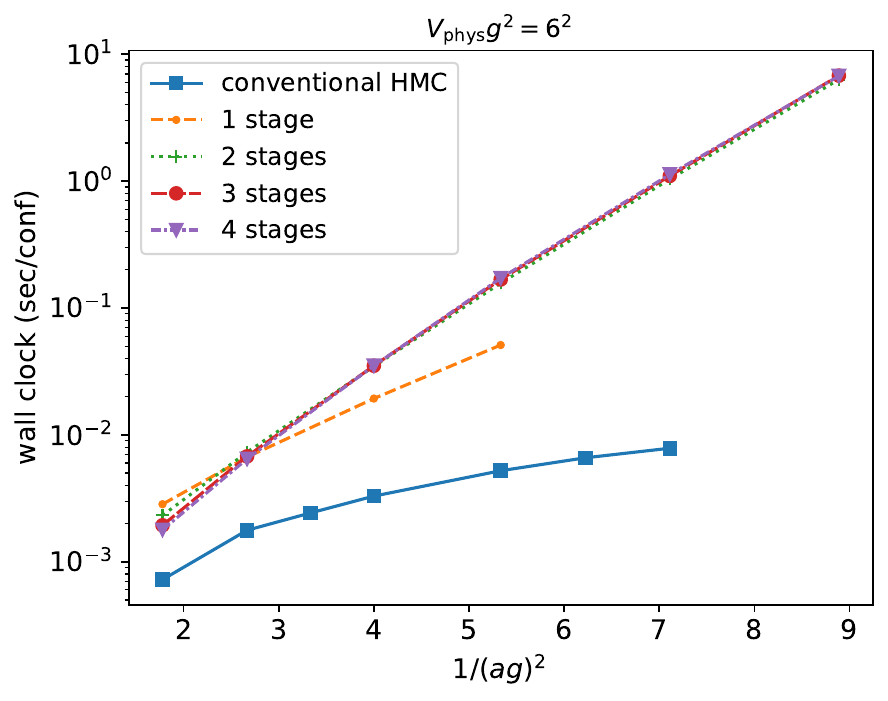}
  \caption{
    The time consumed to generate a configuration
    in wall clock.
    Since the second stage is the most costly,
    this part dominates the time
    even when we include higher stages.
  }
  \label{fig:exe_time}
\end{figure}

\begin{figure}[htb]
  \centering
  \includegraphics[width=75mm]{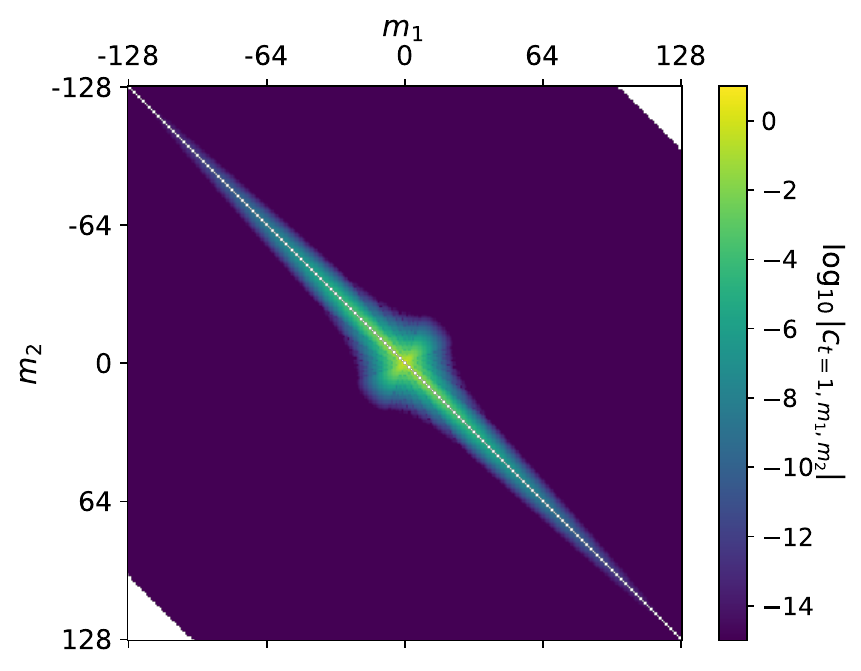}
  \includegraphics[width=75mm]{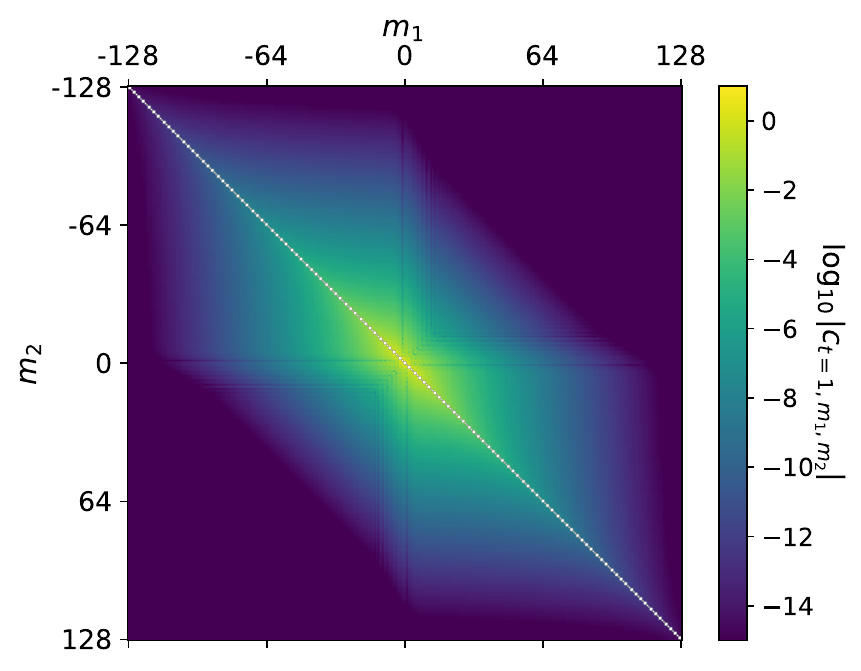}
  \includegraphics[width=75mm]{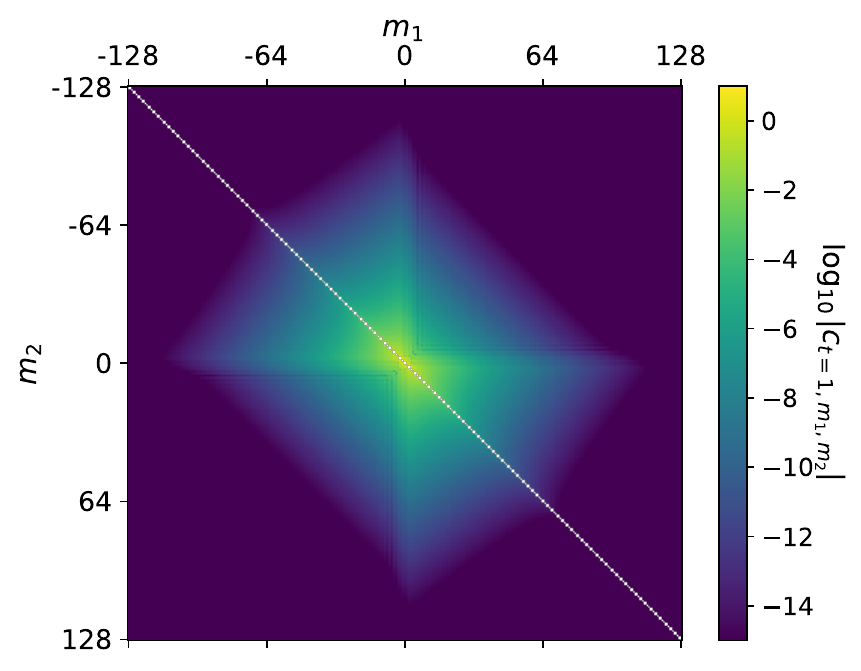}
  \includegraphics[width=75mm]{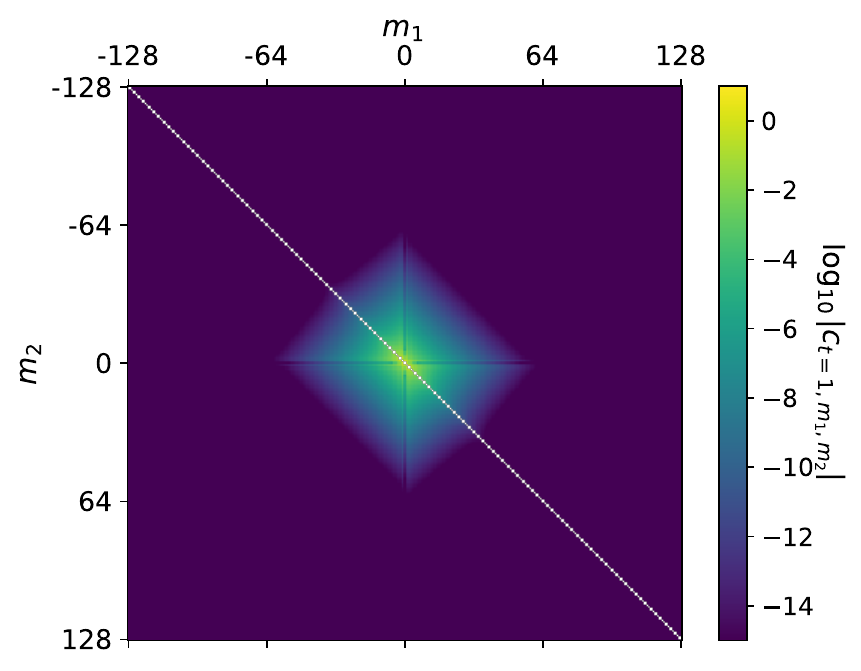}
  \caption{Density plot of the magnitude of the coefficient function in the flow kernel at $t=1$
    (corresponding to the finest lattice during the transformation)
    for $\beta=8.9$.
    The first to fourth stages
    from top left to right bottom.
    The relevant basis functions expand in the $m_1-m_2$ direction after the first stage
    because the effective action includes multiply winding Wilson loops.
    After the second stage, the number of relevant functions decreases
    reflecting that the lattice action is becoming coarse.
  }
  \label{fig:flow_coeffs}
\end{figure}

Concerning the algorithmic overhead,
it is the second stage of decimation
that is the most costly (see figure~\ref{fig:exe_time}).
This is simply because
the number of relevant basis functions is the largest.
Figure~\ref{fig:flow_coeffs}
shows the magnitude of the coefficients
$c_{t,m_1,m_2}$ of eq.~\eqref{eq:flow_kernel}
for $\beta = 8.9$ at $t=1$
(corresponding to the finest lattice during the transformation).
After the first stage,
the functions with large $m_1-m_2$
become relevant as the effective action contains
terms of higher representations, multiply winded loops.
At the later stages, the number of relevant basis functions decreases
reflecting that the lattice is becoming coarse.

\begin{figure}[htb]
  \centering
  \includegraphics[width=75mm]{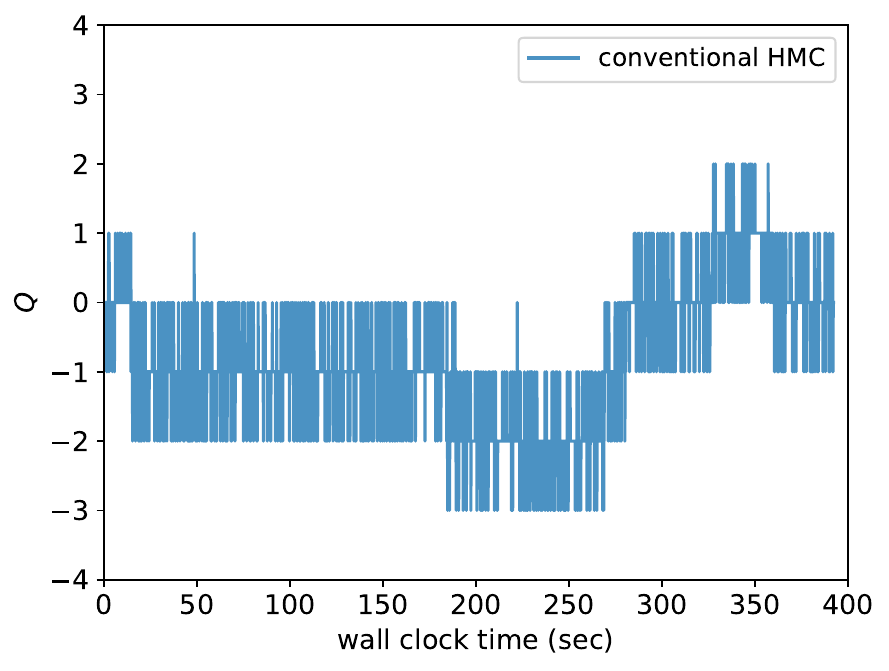}
  \includegraphics[width=75mm]{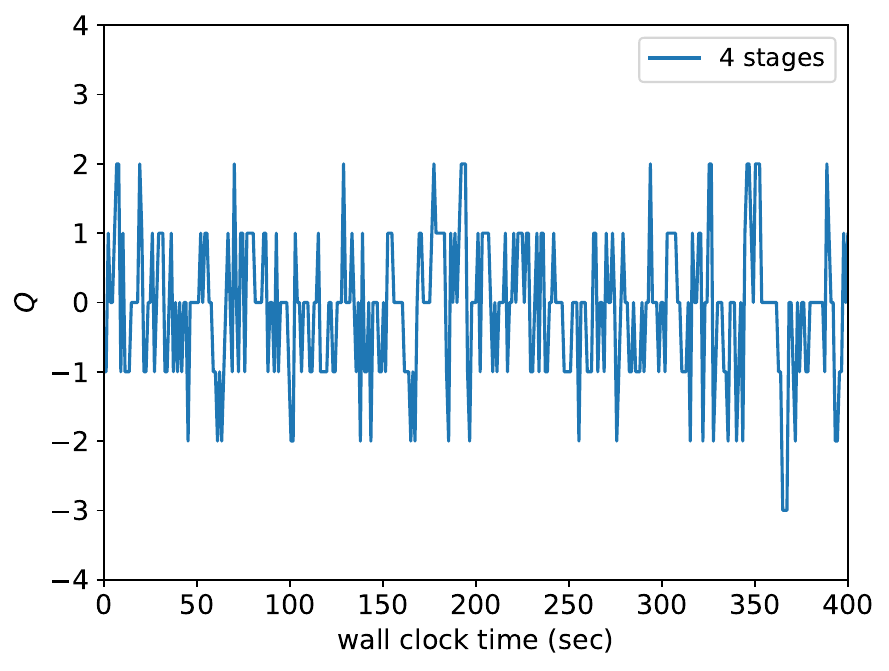}
  \caption{Time series of the topological charge in wall clock
    with the conventional HMC (left)
    and with the decimation map with four stages (right)
    at $\beta=7.1$.
    With the conventional HMC,
    though $Q$ changes back and forth frequently within two nearby sectors,
    it takes a long time to cover all the topological sectors
    because of the large autocorrelation.
    With the decimation map, on the other hand,
    though each update takes a noticeable time,
    it spans all the sectors with a few Monte Carlo steps
    thanks to the minimal autocorrelation.
  }
  \label{fig:time_series}
\end{figure}

Figure~\ref{fig:time_series}
shows the time series of $Q$ at $\beta=7.1$
with and without the decimation map.
With the conventional HMC,
we observe that $Q$ is varying extremely slowly in large scale;
the fast fluctuation only moves $Q$ back and forth between
two nearby sectors.
By implementing the decimation map,
the fluctuation becomes centered correctly at $Q=0$
without visible long-range autocorrelation
even though a single update is taking a noticeable time.

\begin{figure}[htb]
  \centering
  \includegraphics[width=75mm]{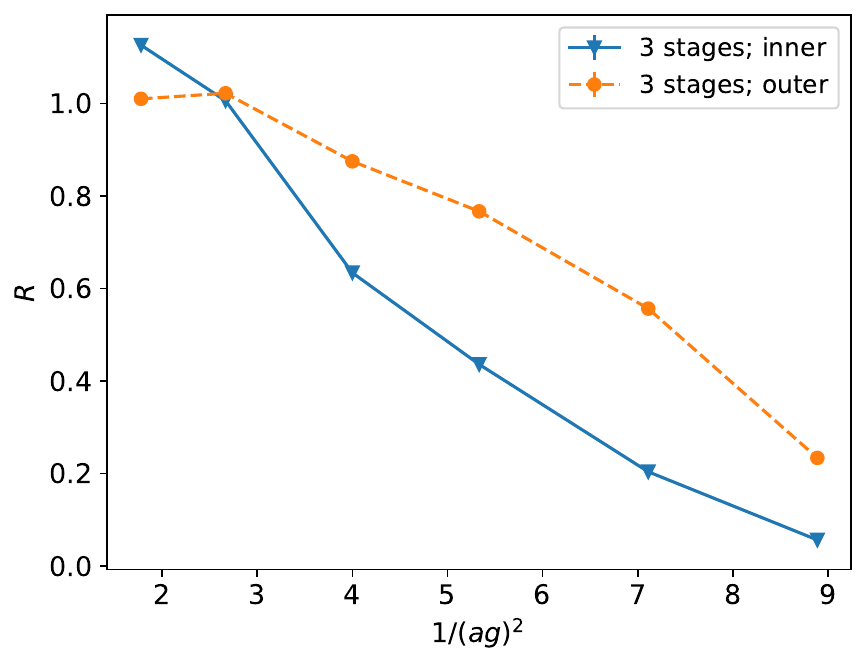}
  \includegraphics[width=75mm]{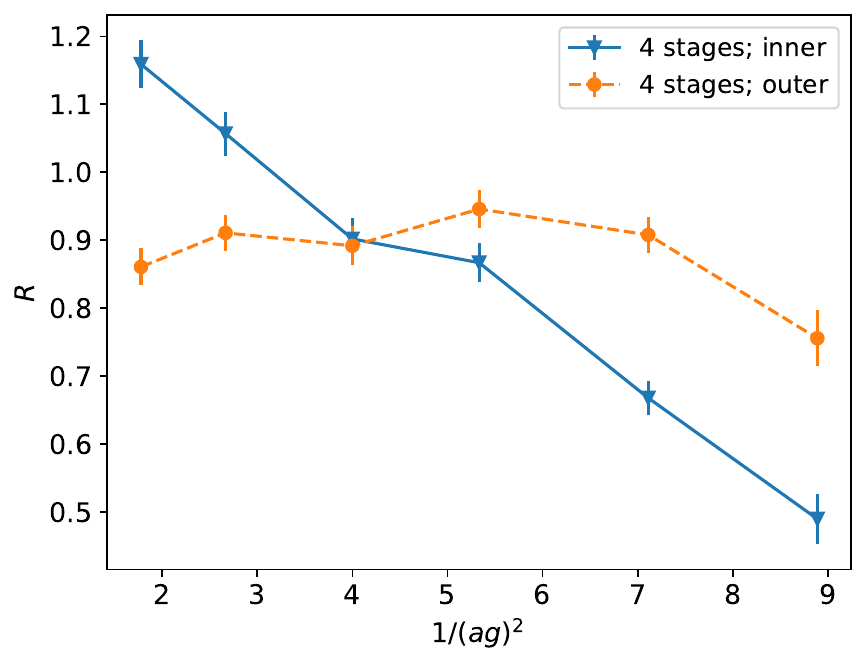}
  \caption{
    Comparison of the tunneling rate $R$
    compared between the inner UV update and
    the outer IR update
    with three and four stages of decimation.
    For small $\beta$,
    since a vortex can fit into a local trivialized region easily,
    the inner update gives larger $R$.
    However,
    as we increase $\beta$,
    the link variables start to form a continuum,
    and consequently they
    need to be updated collectively.
    Accordingly,
    the outer IR update becomes more effective
    at large $\beta$.
  }
  \label{fig:analysis_tunneling}
\end{figure}

Finally, to study
whether it is the UV or IR part of the algorithm
that is generating the acceleration,
we separate the inner (UV) and outer (IR) updates
by freezing one of them alternatively,
and measure the tunneling rate:
\begin{align}
  R \equiv \langle |Q_i - Q_{i+1}| \rangle.
  \label{eq:tunneling_rate}
\end{align}
Figure~\ref{fig:analysis_tunneling} shows
$R$ for each part with various $\beta$.
For small $\beta$, the volume of the trivialized region is
large enough to create a vortex.
Correspondingly,
the inner update has a larger tunneling rate
compared to the outer update.
However, as we enlarge $\beta$,
the chance of creating a vortex inside the trivialized region
becomes small,
and the effectiveness flips.

As discussed in section~\ref{sec:guided_mc},
the outer update is necessary to change the global configuration.
Consequently,
even when the topological charge changes in the UV update,
the central value of the fluctuation will not change
unless the IR modes are altered.
In large $\beta$ theories,
since the problem is to update the IR modes effectively,
mapping the theory to coarse lattice actions
by incorporating renormalization group
seems crucial for the speed up.

\section{Conclusion and Outlook}
\label{sec:conclusion}

We considered the decimation map in the 2D $U(1)$ model
that can be regarded as a coarse-graining transformation.
With this map,
combined with the guided Monte Carlo algorithm,
we showed that the integrated autocorrelation time
of the topological charge can be exponentially improved
in the wall clock time.
We observe x73 speed up
on the $16\times 16$ lattice at $\beta=7.1$
and
the decrease of the exponent by a factor of 0.62.
Our study exemplifies that
it is possible to accelerate the HMC in gauge theories
once an appropriate field transformation is constructed.
We further argued that
the coarse-graining picture is crucial
for accelerating the HMC at large $\beta$.

Towards the application in QCD,
there are three points to be addressed:
generalization to non-Abelian groups,
generalization to higher dimensions,
and inclusion of the fermion;
and none of them is straightforward.
However, machinery of solving the linear equation itself
can be applied to non-Abelian cases in principle
by parameterizing the function space with the Wilson loops.
The nontrivial relations among the Wilson loops
called Mandelstam constraints \cite{Mandelstam:1978ed}
(see also \cite{Boyle:2022xor})
will not be an issue in finding a solution with iterative methods.
Generalization to higher dimension needs work even for the $U(1)$ case
because the variables correlate through various directions
and the effective action involves loops with complicated shapes.
Nevertheless, if we can trivialize codimension-one hypersurfaces
completely,
the resulting effective theory will be described by
the double-sized Wilson loops by gauge invariance.
However, aiming further to include fermion,
it must be necessary to find an approximation scheme
rather than solving the equation exactly
while retaining the renormalization-group picture.
Toy 2D models such as
the $CP^{N-1}$ model and the Schwinger model may be useful for such investigations.
Studies along these lines are in progress and will be reported elsewhere.

\acknowledgments
The authors thank Tom Blum, Peter Boyle, Norman Christ,
Gerald Dunne, Sam Foreman,
Anna Hasenfratz, Luchang Jin, Xiaoyong Jin,
Chulwoo Jung, James Osborn,
Akio Tomiya, Julian Urban
and Urs Wenger for valuable discussions.
This work is supported
by the U.S. Department of Energy under contract No. 
DE-AC02-07CH11359.
N.M. was supported by the Special Postdoctoral Researchers Program of RIKEN.

\end{document}